# Distributed STBCs with Full-diversity Partial Interference Cancellation Decoding

Lakshmi Prasad Natarajan and B. Sundar Rajan
Dept. of ECE, IISc, Bangalore 560012, India
Email: {nlp,bsrajan}@ece.iisc.ernet.in

*Abstract*—Recently, Guo and Xia introduced low complexity decoders called Partial Interference Cancellation (PIC) and PIC with Successive Interference Cancellation (PIC-SIC), which include the Zero Forcing (ZF) and ZF-SIC receivers as special cases, for point-to-point MIMO channels. In this paper, we show that PIC and PIC-SIC decoders are capable of achieving the full cooperative diversity available in wireless relay networks. We give sufficient conditions for a Distributed Space-Time Block Code (DSTBC) to achieve full diversity with PIC and PIC-SIC decoders and construct a new class of DSTBCs with low complexity full-diversity PIC-SIC decoding using complex orthogonal designs. The new class of codes includes a number of known full-diversity PIC/PIC-SIC decodable Space-Time Block Codes (STBCs) constructed for point-to-point channels as special cases. The proposed DSTBCs achieve higher rates (in complex symbols per channel use) than the multigroup ML decodable DSTBCs available in the literature. Simulation results show that the proposed codes have better bit error rate performance than the best known low complexity, full-diversity DSTBCs.

## I. Introduction

In wireless channels, multiple transmit and receive antennas are used to overcome the adverse effects of fading. In systems where the terminals can not have multiple transmit/receive antennas due to space considerations, such as wireless sensor networks or cellular networks for mobile phones, spatial diversity called *cooperative diversity* can be achieved by using the antennas of other users (relays) in the network to aid the communication of messages from a single source [1]. In [2], a two phase amplify and forward (AF) based cooperative protocol was proposed where the relays have no channel state information (CSI) and the destination has full CSI. At the end of the second phase of this protocol, the destination effectively sees a Distributed Space-Time Block Code (DSTBC) being transmitted by the relays. If the number of independent real information symbols in the DSTBC is $K$, then the *rate* of the DSTBC is $R = \frac{K}{2T}$ complex symbols per channel use (cspcu), where $T$ is the combined duration of the first and second phases.

A DSTBC is said to be *g-group maximum-likelihood (ML) decodable* if the $K$ information symbols can be partitioned into $g$ groups, $g > 1$, such that each group of symbols can be ML decoded independent of the symbols of the other groups. If the maximum number of symbols in any group is $\lambda$, then the code is also said to be $\lambda$-*real symbol* or $\frac{\lambda}{2}$-*complex symbol ML decodable*. Multigroup ML decodable DSTBCs based on AF protocol were constructed in [3], [4], [5] and [6]. Note that all these codes depend on the optimal (ML) decoder to tap the full cooperative diversity.

In [7], suboptimal decoders called PIC and PIC-SIC were introduced for decoding STBCs for point-to-point MIMO channels. A PIC decoder partitions the information symbols of the code into multiple groups and decodes each group of symbols independently of other groups. In order to decode a particular group of symbols, a PIC decoder first implements a linear filter to eliminate the interference from symbols in all other groups and then decodes all the symbols of the current group jointly. A PIC-SIC receiver uses successive interference cancellation along with PIC decoding. If $\lambda$ is the maximum number of symbols in any group, we say that the PIC or PIC-SIC decoder performs $\lambda$-*real symbol PIC* or *PIC-SIC decoding* respectively. When $\lambda = 1$, the PIC (PIC-SIC) decoder reduces to ZF (ZF-SIC) decoder. Sufficient conditions for an STBC to achieve full-diversity in point-to-point MIMO channel with PIC and PIC-SIC decoding were given in [8], [9]. Code constructions for point-to-point MIMO channels with full-diversity and low complexity PIC/PIC-SIC decoding were given in [9], [10], [11], [12] and with ZF/ZF-SIC decoding were given in [13], [14]. For both multigroup ML decoding and PIC/PIC-SIC decoding, the complexity of the decoder is determined by the number of symbols per group $\lambda$.

The contributions and organization of this paper are as follows.

- We show that PIC and PIC-SIC decoders are capable of achieving the full cooperative diversity offered by the wireless relay network. For a two phase amplify and forward based cooperative protocol, we give sufficient conditions for a DSTBC to achieve full cooperative diversity when PIC and PIC-SIC decoders are used at the destination. As a special case, we also obtain full-diversity criteria for ZF and ZF-SIC decoding (Section III).
- We construct a new class of full-diversity PIC/PIC-SIC decodable DSTBCs using Complex Orthogonal Designs [15]. We then identify a subclass of the new family of codes that contains $\lambda$-real symbol PIC-SIC decodable DSTBCs for any number of relays $N \geq 1$ and $\lambda \leq N$ with rates arbitrarily close to $\frac{\lambda}{\lambda+1}$ cspcu. The new class of codes includes a number of known full-diversity PIC/PIC-SIC decodable STBCs constructed for point-to-point channels as special cases, such as the codes in [9], [10], [12], a family of codes in [11] and the Toeplitz codes [13] (Section IV).

- The proposed full-diversity DSTBCs achieve higher rates when compared with the known multigroup ML decodable DSTBCs of similar decoding complexity (see Table I in Section V). We also present simulation results which show that the new PIC-SIC decodable codes have a better bit error rate performance than the best known multigroup ML decodable DSTBCs (Section V).

The system model is explained in Section II and some related open problems are discussed in Section VI.

**Notation:** For a complex matrix $A$ the transpose, the conjugate and the conjugate-transpose are denoted by $A^T$, $A^*$ and $A^H$ respectively, $||A||_F$ is the Frobenius norm of the matrix $A$, $I_n$ is the $n \times n$ identity matrix, $\mathbf{0}$ is the all zero matrix of appropriate dimension and $i = \sqrt{-1}$. The cardinality of a set $\Gamma$ is denoted by $|\Gamma|$. The complement of a set $\Gamma$ with respect to a universal set $U$ is denoted by $\Gamma^c$, whenever $U$ is clear from context. For a complex matrix $A$, $A_{Re}$ and $A_{Im}$ are its real and imaginary parts respectively and $vec(A)$ is the vectorization of $A$. The expectation operator is denoted by $\mathsf{E}(\cdot)$.

## II. SYSTEM MODEL

We consider a wireless relay network with a source node, $N$ relay nodes and a destination node. The source and the relay nodes are equipped with single antennas and the destination has $N_D$ antennas as shown in Fig. 1. The channel gain from the source to the $j^{th}$ relay is $f_j$, and the channel gain from the $j^{th}$ relay to the $l^{th}$ receive antenna at the destination is $g_{j,l}$, for $j = 1, \ldots, N$ and $l = 1, \ldots, N_D$. We make the following assumptions: (*i*) All the nodes are half-duplex constrained, (*ii*) the channel gains $f_j$ and $g_{j,l}$, $j = 1, \ldots, N$, $l = 1, \ldots, N_D$ are independent circularly symmetric complex Gaussian random variables with zero mean and unit variance and with coherence interval of duration at least $T_1$ and $T_2$ respectively, (*iii*) the relay nodes have no channel state information and the destination has the knowledge of all channel gains $f_j$, $g_{j,l}$, and (*iv*) the transmissions from the relay nodes to the destination are synchronized at the symbol level.

In each transmission cycle, the source transmits $K$ real information symbols $x_1, \ldots, x_K$. The source is equipped with a finite subset $\mathcal{A} \subset \mathbb{R}^K$ called the *signal set* and $K$ complex vectors $\{\nu_1, \ldots, \nu_K\} \subset \mathbb{C}^{T_1}$, that are linearly independent over $\mathbb{R}$. The information vector $x = [x_1, \ldots, x_K]^T$ assumes values from $\mathcal{A}$. During a transmission cycle, say the information symbol vector assumes the value $[a_1, \ldots, a_K]^T \in \mathcal{A}$. During the *broadcast phase*, the source synthesizes the vector $z = \sum_{i=1}^{K} a_i \nu_i \in \mathbb{C}^{T_1}$ and transmits $\sqrt{\pi_1 P} z$ to all the relays, where $P$ is the average power transmitted in the network and $\pi_1 > 0$. The signal set $\mathcal{A}$ and the vectors $\nu_i$ are chosen in such a way that $\mathsf{E}\left(||z||_F^2\right) = T_1$. The vector received by the $j^{th}$ relay is $r_j = f_j \sqrt{\pi_1 P} z + v_j$, $j = 1, \ldots, N$. Here, $v_j$ is the additive white Gaussian noise vector at the $j^{th}$ relay and it has zero mean and covariance $I_{T_1}$. In the *cooperation phase*, the $j^{th}$ relay transmits a linearly processed version of either $r_j$ or $r_j^*$. The subset $\mathcal{S} \subset \{1, \ldots, N\}$ denotes the set of indices of the relays that process $r_j^*$. For any indexed set of $N$ matrices or scalars $\{C_1, \ldots, C_N\}$, let $\overline{C}_j = C_j^*$ if $j \in \mathcal{S}$, and $\overline{C}_j = C_j$ else. The $j^{th}$ relay is equipped with a matrix $B_j \in \mathbb{C}^{T_2 \times T_1}$. In the cooperation phase, the $j^{th}$ relay transmits $t_j = \sqrt{\frac{\pi_2 P}{\pi_1 P+1}} \overline{B_j r_j}$
$= \sqrt{\frac{\pi_1 \pi_2 P^2}{\pi_1 P+1}} \overline{f_j} \overline{B_j z} + \sqrt{\frac{\pi_2 P}{\pi_1 P+1}} \overline{B_j v_j}$. The real numbers $\pi_1, \pi_2 > 0$ are chosen such that $\pi_1 T_1 + \pi_2 R T_2 = T_1 + T_2$. The signal received by the $l^{th}$ antenna, $l = 1, \ldots, N_D$, at the destination during cooperation phase is $y_l = \sum_{j=1}^{N} g_{j,l} t_j + w_l$
$= \sum_{j=1}^{N} \left( g_{j,l} \overline{f_j} \sqrt{\frac{\pi_1 \pi_2 P^2}{\pi_1 P+1}} \overline{B_j z} + \sqrt{\frac{\pi_2 P}{\pi_1 P+1}} g_{j,l} \overline{B_j v_j} \right) + w_l$.
Here, $w_l$ is the additive circularly symmetric complex Gaussian noise at the $l^{th}$ receive antenna with zero mean and covariance $I_{T_2}$. The $T \times N_D$ received matrix $Y = [y_1 \ y_2 \cdots y_{N_D}]$ satisfies $Y = \sqrt{\frac{\pi_1 \pi_2 P^2}{\pi_1 P+1}} XH + U$, where $X = [\overline{B_1 z} \ \overline{B_2 z} \cdots \overline{B_N z}] \in \mathbb{C}^{T_2 \times N}$ is the codeword matrix, $H = F\mathcal{G}$ is the channel matrix with $F = diag(\overline{f_1}, \ldots, \overline{f_N})$ and the $(j,l)^{th}$ entry of the matrix $\mathcal{G} \in \mathbb{C}^{N \times N_D}$ being $g_{j,l}$. The matrix $U \in \mathbb{C}^{T_2 \times N_D}$ is the total noise seen by the receiver. If we denote the columns of $U$ by $u_l$, $l = 1, \ldots, N_D$, then

$$u_l = \sum_{j=1}^{N} \sqrt{\frac{\pi_2 P}{\pi_1 P+1}} g_{j,l} \overline{B_j v_j} + w_l. \quad (1)$$

The noise vector $vec(U) = [u_1^T, u_2^T, \ldots, u_{N_D}^T]^T$ is zero mean circularly symmetric complex Gaussian. Since $z = \sum_{i=1}^{K} a_i \nu_i$, it is clear that each entry of the codeword matrix $X = [\overline{B_1 z} \ \overline{B_2 z} \cdots \overline{B_N z}]$ is a complex linear combination of $a_i$, $i = 1, \ldots, K$. Thus, there exist matrices $A_i \in \mathbb{C}^{T_2 \times N}$, $i = 1, \ldots, K$, such that the set of codewords is $\mathcal{C} = \{\sum_{i=1}^{K} a_i A_i | [a_1, \ldots, a_K]^T \in \mathcal{A}\}$. The matrices $A_i$ are the *linear dispersion* or *weight matrices*. The finite set of matrices $\mathcal{C}$ is the DSTBC and the underlying design is $\mathbf{X} = \sum_{i=1}^{K} x_i A_i$. The rate of the DSTBC $\mathcal{C}$ in complex symbols per channel use is $R = \frac{K}{2(T_1+T_2)}$, and in bits per channel use is $\frac{\log_2 |\mathcal{A}|}{T_1+T_2}$. Note that each column of the codeword matrix $X$ is either a linear transformation of the vector $z$ or its conjugate $z^*$. Such codes are said to be *conjugate linear* [6].

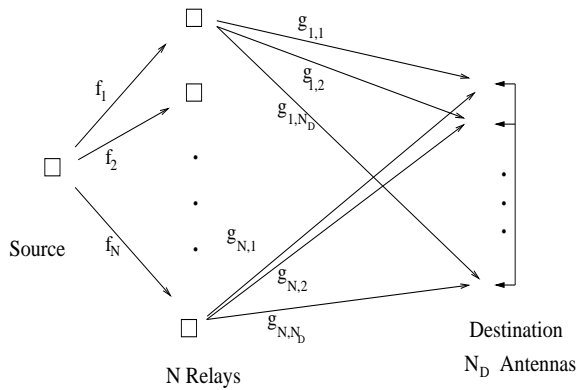

Fig. 1. Relay network model

## III. PARTIAL INTERFERENCE CANCELLATION DECODING AND FULL-DIVERSITY CRITERION

### A. PIC and PIC-SIC decoding of DSTBCs

Consider a DSTBC in $K$ real symbols. A *grouping scheme* [7] is a partition $\mathcal{I}_1, \ldots, \mathcal{I}_g$ of the set $\{1, \ldots, K\}$, where $\mathcal{I}_k$ are called *groups*. There is a corresponding partition of the information symbols into $g$ vectors, where for $k = 1, \ldots, g$, the $k^{th}$ vector of information symbols is $x_{\mathcal{I}_k} = [x_{i_{k,1}}, x_{i_{k,2}}, \ldots, x_{i_{k,|\mathcal{I}_k|}}]^T$, where $\mathcal{I}_k = \{i_{k,1}, i_{k,2}, \ldots, i_{k,|\mathcal{I}_k|}\}$ with $i_{k,1} < i_{k,2} < \cdots < i_{k,|\mathcal{I}_k|}$. Let the $g$ groups of information symbols be encoded independently of each other, i.e., the DSTBC

$$\mathcal{C} = \left\{ \sum_{i=1}^{K} x_i A_i \middle| x_{\mathcal{I}_k} \in \mathcal{A}_{\mathcal{I}_k},\ k = 1, \ldots, g \right\} \subset \mathbb{C}^{T_2 \times N}, \quad (2)$$

for some finite subsets $\mathcal{A}_{\mathcal{I}_k} \subset \mathbb{R}^{|\mathcal{I}_k|}$, $k = 1, \ldots, g$. For a complex matrix $A$, let $\widetilde{vec}(A) = [vec(A_{Re})^T\ vec(A_{Im})^T]^T$. The received matrix $Y$ can be rewritten as $y' = \widetilde{vec}(Y) = \sqrt{\frac{\pi_1 \pi_2 P^2}{\pi_1 P + 1}} \sum_{i=1}^{K} x_i \widetilde{vec}(A_i H) + \widetilde{vec}(U) = G'x + \widetilde{vec}(U)$, where $G' =$

$$\sqrt{\frac{\pi_1 \pi_2 P^2}{\pi_1 P + 1}} [\widetilde{vec}(A_1 H) \cdots \widetilde{vec}(A_K H)] \in \mathbb{R}^{2N_D T_2 \times K}, \quad (3)$$

and $x = [x_1, x_2, \ldots, x_K]^T$. Consider $y = \Gamma^{-\frac{1}{2}} y' = Gx + n$, where, $\Gamma$ is the covariance of $\widetilde{vec}(U)$, $G = \Gamma^{-\frac{1}{2}} G'$ and $n = \Gamma^{-\frac{1}{2}} \widetilde{vec}(U)$ is a zero mean real Gaussian vector with covariance $I_{2N_D T_2}$. Let $G = [g_1\ g_2 \cdots g_K]$, where $g_i$, $i = 1, \ldots, K$, are the column vectors of $G$. For any nonempty subset $\mathcal{I} = \{i_1, \ldots, i_{|\mathcal{I}|}\} \subset \{1, \ldots, K\}$, with $i_1 < i_2 < \cdots < i_{|\mathcal{I}|}$, let $G_{\mathcal{I}} = [g_{i_1}\ g_{i_2} \cdots g_{i_{|\mathcal{I}|}}]$. Let $V_{\mathcal{I}_k}$ be the column space of the matrix $G_{\mathcal{I}_k^c}$ and $P_{\mathcal{I}_k}$ be the matrix that projects a vector onto the subspace $V_{\mathcal{I}_k}^{\perp}$, the orthogonal complement of the subspace $V_{\mathcal{I}_k}$. Also, let $\tilde{\mathcal{I}}_k = \cup_{\ell > k} \mathcal{I}_\ell$, $\tilde{V}_{\mathcal{I}_k}$ be the column space of the matrix $G_{\tilde{\mathcal{I}}_k}$ and $\tilde{P}_{\mathcal{I}_k}$ be the matrix that projects a vector onto the subspace $\tilde{V}_{\mathcal{I}_k}^{\perp}$. The PIC decoding of the DSTBC is performed as follows for $k = 1, \ldots, g$,

$$\hat{x}_{\mathcal{I}_k} := arg\ min_{x_{\mathcal{I}_k} \in \mathcal{A}_{\mathcal{I}_k}} \|P_{\mathcal{I}_k} y - P_{\mathcal{I}_k} G_{\mathcal{I}_k} x_{\mathcal{I}_k}\|_F^2. \quad (4)$$

The PIC-SIC decoding of the DSTBC is performed as given by the following algorithm. The decoder is initialized with $k = 1$ and $y_1 = y$.

- Step 1: Decode the $k^{th}$ vector of information symbols as

$$\hat{x}_{\mathcal{I}_k} := arg\ min_{x_{\mathcal{I}_k} \in \mathcal{A}_{\mathcal{I}_k}} \|\tilde{P}_{\mathcal{I}_k} y_k - \tilde{P}_{\mathcal{I}_k} G_{\mathcal{I}_k} x_{\mathcal{I}_k}\|_F^2.$$

- Step 2: Assign $y_{k+1} := y_k - G_{\mathcal{I}_k} \hat{x}_{\mathcal{I}_k}$ and then $k := k+1$.
- Step 3: If $k > g$, stop. Else, go to Step 1.

### B. Full-diversity criteria

Let $\mathcal{I} = \{i_1, \ldots, i_{|\Gamma|}\}$ be any non-empty subset of $\{1, \ldots, K\}$ with $i_1 < i_2 < \cdots < i_{|\Gamma|}$. For any $u = [u_1, \ldots, u_{|\mathcal{I}|}]^T \in \mathbb{R}^{|\mathcal{I}|}$, define $X_{\mathcal{I}}(u) = \sum_{j=1}^{|\mathcal{I}|} u_i A_{i_j}$. For any set of vectors $\mathcal{A}$, let $\Delta \mathcal{A} = \{a_1 - a_2 | a_1, a_2 \in \mathcal{A}\}$.

*Theorem 1:* The PIC decoding of the DSTBC $\mathcal{C}$ in (2) with the grouping scheme $\mathcal{I}_1, \ldots, \mathcal{I}_g$ achieves a diversity of $N \left(1 - \frac{log(log P)}{log P}\right)$ for $N_D = 1$ and a diversity of $N$ for $N_D > 1$, if the following condition is satisfied for every $k = 1, \ldots, g$:

- for every $a_k \in \Delta \mathcal{A}_{\mathcal{I}_k} \setminus \{0\}$ and every $u \in \mathbb{R}^{|\mathcal{I}_k^c|}$, the rank of $\left(X_{\mathcal{I}_k}(a_k) + X_{\mathcal{I}_k^c}(u)\right)$ is $N$.

*Proof:* Proof is given in Appendix A. ∎

*Theorem 2:* The PIC-SIC decoding of the DSTBC $\mathcal{C}$ in (2) with the grouping scheme $\mathcal{I}_1, \ldots, \mathcal{I}_g$ achieves a diversity of $N \left(1 - \frac{log(log P)}{log P}\right)$ if $N_D = 1$ and a diversity of $N$ if $N_D > 1$, if the following condition is satisfied for every $k = 1, \ldots, g$:

- for every $a_k \in \Delta \mathcal{A}_{\mathcal{I}_k} \setminus \{0\}$ and every $u \in \mathbb{R}^{|\tilde{\mathcal{I}}_k|}$, the rank of $\left(X_{\mathcal{I}_k}(a_k) + X_{\tilde{\mathcal{I}}_k}(u)\right)$ is $N$.

*Proof:* The proof is similar to that of Theorem 1. We give an outline of the proof in Appendix B. ∎

The class of PIC and PIC-SIC decoders contains the ZF and ZF-SIC decoders as special cases. When each real symbol $x_i$, $i = 1, \ldots, K$, forms a group by itself, the PIC decoder reduces to the ZF decoder and the PIC-SIC decoder reduces to the ZF-SIC decoder.

*Corollary 1:* The DSTBC $\mathcal{C}$ in (2) achieves a diversity of $N \left(1 - \frac{log(log P)}{log P}\right)$ for $N_D = 1$ and a diversity of $N$ for $N_D > 1$, with ZF decoding and ZF-SIC decoding with any ordering, if the rank of $\sum_{i=1}^{K} u_i A_i$ is $N$ for every $u = [u_1, \ldots, u_K]^T \in \mathbb{R}^K \setminus \{0\}$.

*Proof:* It is straightforward to show that the criteria of Theorems 1 and 2 are satisfied for the grouping scheme corresponding to ZF and ZF-SIC decoders under the hypothesis of this theorem. ∎

The diversity promised by Theorems 1 and 2 is equal to the full cooperative diversity obtainable by using the optimal i.e., ML decoder at the destination [2]. We thus say that the DSTBCs satisfying the conditions in Theorems 1 and 2 achieve *full-diversity* under PIC and PIC-SIC decoding respectively. In Appendix C we show that codes constructed using the new full-diversity criteria are resistant to relay node failures as well.

The criteria in Theorems 1 and 2 are the same as the criteria given in [9] for an STBC to achieve full-diversity in a point-to-point MIMO channel with PIC/PIC-SIC decoding. Further, as shown in [9], these are equivalent to the criteria given in [8] for achieving full-diversity in a point-to-point MIMO channel. In this case $\mathcal{C}$ in (2) is an STBC for a MIMO channel with $N$ transmit antennas and a delay of $T_2$ channel uses. Thus, all known full-diversity PIC/PIC-SIC decodable, conjugate linear codes designed for the collocated MIMO channel can be used as DSTBCs to achieve full-diversity in a relay network with PIC/PIC-SIC decoding. For example, the Overlapped Alamouti Codes [14] are conjugate linear and are known to achieve full diversity in point-to-point MIMO channels with ZF and ZF-SIC decoding, and thus, they are full-diversity ZF/ZF-SIC decodable DSTBCs as well.

## IV. A NEW CLASS OF DSTBCS WITH FULL-DIVERSITY PIC/PIC-SIC DECODING

Let the number of relays be $N$ and $N', L > 0$ be integers such that $N = LN'$. Let $\lambda \leq L$ be a positive integer. We construct full-diversity PIC/PIC-SIC decodable codes for $N$ relays with $\lambda$ symbols per decoding group, using a Complex Orthogonal Design (COD) for $N'$ antennas. Let $\mathbf{W}$ be a $T' \times N'$ COD in $K'$ real symbols. The DSTBCs in the proposed class are parametrized by the tuple $(N, \mathbf{W}, \lambda, n)$, where $n \geq 1$ is an integer. The construction for a given $(N, \mathbf{W}, \lambda, n)$ is as follows.

The number of symbols in the resulting design $\mathbf{X}_{N,\mathbf{W},\lambda,n}$ is $K = \lambda n K'$ and the number of decoding groups is $g = nK'$. For $k = 1, \ldots, g$, the $k^{th}$ group is

$$\mathcal{I}_k = \{(k-1)\lambda + 1, (k-1)\lambda + 2, \ldots, k\lambda\}, \quad (5)$$

i.e., the first $\lambda$ symbols $x_1, \ldots, x_\lambda$ form the first group, the second $\lambda$ symbols $x_{\lambda+1}, \ldots, x_{2\lambda}$ form the second group and so on. For $m \in \{1, \ldots, n\}$ and $\ell \in \{1, \ldots, \lambda\}$, define $\mathbf{W}(m, \ell)$ to be the $T' \times N'$ COD $\mathbf{W}$ in the $K'$ real symbols $x_{\lambda K'(m-1)+\ell}, x_{\lambda K'(m-1)+\ell+\lambda}, \ldots, x_{\lambda K'(m-1)+\ell+\lambda(K'-1)}$.
From (5), we know that for $m \in \{1, \ldots, n\}$ and $i \in \{1, \ldots, K'\}$, the symbols $x_{\lambda K'(m-1)+1+(i-1)\lambda}, x_{\lambda K'(m-1)+2+(i-1)\lambda}, \ldots, x_{\lambda K'(m-1)+\lambda+(i-1)\lambda}$ form the $(K'(m-1)+i)^{th}$ group. Thus, the $i^{th}$ symbol of each of the designs $\mathbf{W}(m,1), \mathbf{W}(m,2), \ldots, \mathbf{W}(m,\lambda)$ together form the $(K'(m-1)+i)^{th}$ group. Now, for $m \in \{1, \ldots, n\}$ and $\ell = \lambda + 1, \ldots, L$ define $\mathbf{W}(m, \ell)$ recursively as $\mathbf{W}(m, \ell) = \mathbf{W}(m, \ell - \lambda)$. The proposed design is

$$\mathbf{X}_{N,\mathbf{W},\lambda,n} = \begin{bmatrix} \mathbf{W}(1,1) & \mathbf{0} & \cdots & \mathbf{0} \\ \mathbf{W}(2,1) & \mathbf{W}(1,2) & \cdots & \vdots \\ \vdots & \mathbf{W}(2,2) & \ddots & \mathbf{0} \\ \vdots & \vdots & \ddots & \mathbf{W}(1,L) \\ \vdots & \vdots & \cdots & \mathbf{W}(2,L) \\ \vdots & \vdots & \cdots & \vdots \\ \mathbf{W}(n,1) & \vdots & \cdots & \vdots \\ \mathbf{0} & \mathbf{W}(n,2) & \cdots & \vdots \\ \vdots & \vdots & \ddots & \vdots \\ \mathbf{0} & \mathbf{0} & \cdots & \mathbf{W}(n,L) \end{bmatrix}, \quad (6)$$

where each $\mathbf{0}$ is a $T' \times N'$ all zero matrix. The design (6) consists of $n$ diagonal layers. The $m^{th}$ diagonal layer encodes the $K'$ groups $\mathcal{I}_{K'(m-1)+1}, \mathcal{I}_{K'(m-1)+2}, \ldots, \mathcal{I}_{K'm}$. For $k = 1, \ldots, g$, the symbol vector $x_{\mathcal{I}_k}$ is encoded using a finite subset of $\mathbb{Z}^\lambda$ rotated by a full-diversity rotation matrix [16], $Q \in \mathbb{R}^{\lambda \times \lambda}$, i.e., $\mathcal{A}_{\mathcal{I}_k} \subset Q\mathbb{Z}^\lambda$. The delay of $\mathbf{X}_{N,\mathbf{W},\lambda,n}$ is $T_2 = (n + L - 1)T'$.

*Proposition 1:* The subclass of DSTBCs proposed in this section corresponding to $\lambda = 1$ give full diversity with ZF decoding and ZF-SIC decoding under any successive interference cancellation ordering.

*Proof:* See Appendix D. ∎

*Proposition 2:* All the DSTBCs proposed in this section give full-diversity with PIC-SIC decoding under the grouping scheme in (5). Further, the subclass corresponding to $n = 1, 2$ give full-diversity with PIC decoding under the grouping scheme in (5).

*Proof:* See Appendix E. ∎

*Example 1: A new family of full-diversity PIC-SIC decodable DSTBCs based on the Alamouti design:* Consider the subclass of proposed codes with $\mathbf{W}$ as the Alamouti design $\begin{bmatrix} w_1 & w_2 \\ -w_2^* & w_1^* \end{bmatrix}$. These codes have parameters $N' = T' = 2$, $N$ an even positive integer, $L = \frac{N}{2}$, $1 \leq \lambda \leq \frac{N}{2}$, $T_2 = N + 2(n-1)$ and $K = 4n\lambda$. Let $\mathbf{D}_\ell = [\mathbf{W}(1, \ell)^T, \mathbf{W}(2, \ell)^T, \ldots, \mathbf{W}(n, \ell)^T]^T$ for $1 \leq \ell \leq \lambda$ and let $\mathbf{D} = [\mathbf{D}_1^T, \mathbf{D}_2^T, \ldots, \mathbf{D}_\lambda^T]^T$. The design $\mathbf{D}$ contains $n\lambda$ Alamouti blocks, placed one below the other. Because of the Alamouti structure, the second column of $\mathbf{D}$ is composed of complex variables that are conjugates of the complex variables $x_p \pm ix_q$ appearing in the first column of $\mathbf{D}$. Further, the first column of $\mathbf{D}$ contains all the $2n\lambda$ complex symbols $x_p \pm ix_q$ appearing as entries in the design (6). Note that all the entries appearing in the odd columns of (6) are contained in the first column of $\mathbf{D}$ and all the entries in the even columns of (6) are contained in the second column of $\mathbf{D}$. If we choose $z$ as the first column of $\mathbf{D}$, then the $j^{th}$ column of the design (6), $j = 1, 3, \ldots, N - 1$, can be expressed as $B_j z$ for some $B_j \in \mathbb{C}^{T_2 \times 2n\lambda}$. Similarly, the $j^{th}$ column for $j = 2, 4, \ldots, N$ can be expressed as $B_j^* z^*$ for some $B_j \in \mathbb{C}^{T_2 \times 2n\lambda}$. Thus, the resulting design (6) is conjugate linear and $T_1 = 2n\lambda$ is the length of the vector $z$ that the source transmits to the relays during the broadcast phase. The rate of the DSTBC is $R = \frac{\lambda}{\lambda + 1 + \frac{N-2}{2n}}$. By increasing $n$, rates arbitrarily close to $\frac{\lambda}{\lambda+1}$ can be achieved.

*Example 2: A family of codes from [9]:* In Appendix F we show that a family of codes from [9] constructed in the context of point-to-point MIMO is a special case of our construction procedure. The resulting DSTBCs include codes for all $N \geq 1$, $\lambda \leq N$ with rates arbitrarily close to $\frac{\lambda}{\lambda+1}$.

For equal values of $N$, $T_2$ and $\lambda$, the codes of Example 1 have slightly higher rate than the codes of Example 2. In Appendix F, we show that the following full-diversity PIC/PIC-SIC decodable STBCs constructed for point-to-point MIMO channel are specific examples of the new class of codes: codes in [10], [12], the Toeplitz codes [13] and a family of codes from [11].

## V. COMPARISON WITH MULTIGROUP ML DECODABLE FULL-DIVERSITY DSTBCS

### A. Comparison of achievable rates

Table I summarizes the comparison of achievable rates of the new codes with known low decoding complexity DSTBCs. Single real symbol ML decodable DSTBCs were constructed in [3] with rate at the most $\frac{2}{2+N}$, and in [4] with rate $\frac{1}{4}$. The new codes of Example 1 and the codes in Example 2 corresponding to $\lambda = 1$ can achieve rates upto $\frac{1}{2}$ cspcu, which is twice the maximum rate reported in [4]. Single complex symbol ML decodable codes for any number of relays $N \geq 4$ were constructed in [5] with rate at the most $\frac{4}{4+N}$ (which, in turn, is upper bounded by $\frac{1}{2}$). The $\lambda = 2$ codes of Example 1 achieve rates upto $\frac{2}{3}$ irrespective of the number of relays. In [6], 4-group ML decodable DSTBCs were constructed for even number of relays with rate $\frac{1}{2}$ cspcu, corresponding to $\lambda = \frac{N}{2}$. For $\lambda = \frac{N}{2}$, the codes in Examples 1 and 2 can achieve rates arbitrarily close to $\frac{N}{N+2}$, which is higher than $\frac{1}{2}$ for $N > 2$ and equals $\frac{1}{2}$ for $N = 2$.

TABLE I
COMPARISON OF FULL-DIVERSITY, LOW DECODING COMPLEXITY DSTBCS

|  | Yi et. al. [3] | Srinath et. al [4] | Harshan et. al. [5] | Rajan et. al. [6] | **Codes in Example 1** | **Codes in Example 2** |
|---|---|---|---|---|---|---|
| Number of relays, $N$ | $\geq 2$ | $\geq 1$ | $\geq 4$ | $2m, m \geq 1$ | **$2m, m \geq 1$** | $\geq 1$ |
| Real symbols per group, $\lambda$ | 1 | 1 | 2 | $\frac{N}{2}$ | $\leq \frac{N}{2}$ | $\leq N$ |
| Rate, $R$ | $\frac{2}{2+N}$ † | $\frac{1}{4}$ | $\frac{4}{4+N}$ † | $\frac{1}{2}$ | $\frac{\lambda}{\lambda+1}$ ‡ | $\frac{\lambda}{\lambda+1}$ ‡ |
| Full-diversity decoding method | ML | ML | ML | ML | **PIC-SIC** | **PIC-SIC** |

† Upper bound on achievable rate. ‡Supremum of achievable rates.

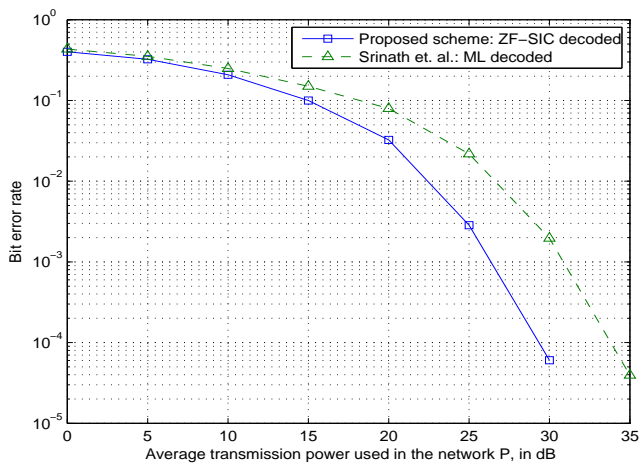

Fig. 2. BER comparison: $\lambda = 1$, Relay network $N = 8$, $N_D = 1$, 2 bpcu.

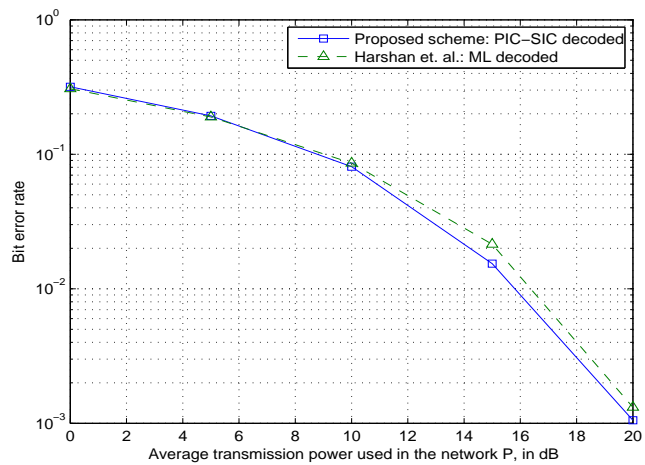

Fig. 3. BER comparison: $\lambda = 2$, Relay network $N = 6$, $N_D = 4$, 2 bpcu.

*B. Simulation Results*

For all the codes we use Gray mapping to convert bits to information symbols and the power allocation $\pi_1 = 1$ and $\pi_2 = \frac{1}{R}$. This is the optimal power allocation when the destination uses an ML decoder to decode a DSTBC [2]. In Fig. 2, we compare single real symbol decodable DSTBCs for $N = 8$, $N_D = 1$ and 2 bits per channel use (bpcu). The new code from Example 1 uses $n = 3$, has a symbol rate of $\frac{1}{3}$ cspcu, uses 8-PAM alphabet and is ZF-SIC decoded. It is compared with the rate $\frac{1}{4}$ cspcu code from Srinath et. al. [4]. The code from [4] uses 16-PAM and is ML decoded. The new code outperforms the code from [4] by a large margin. In Fig. 3, we compare single complex symbol decodable DSTBCs, i.e., those with $\lambda = 2$ for $N = 6$ relays and $N_D = 4$ antennas at the destination at a bit rate of 2 bpcu. The new code of Example 1 with $n = 2$, i.e., $R = \frac{1}{2}$ cspcu is compared with the code from Harshan et. al. [5]. The new code uses rotated 16-QAM and is PIC-SIC decoded. The code from [5] has $R = \frac{1}{3}$ only, is encoded with rotated 64-QAM and is ML decoded. The figure shows that the new code outperforms the code from [5].

## VI. DISCUSSION

In this paper, we have shown that PIC and PIC-SIC decoders are capable of achieving full cooperative diversity in two hop wireless relay networks where the source is employed with a single antenna. We have constructed a new class of full-diversity PIC/PIC-SIC decodable DSTBCs that includes most known full-diversity PIC/PIC-SIC decodable STBCs as special cases. We have also shown that these new codes achieve higher rates and better bit error rate performance than known multigroup ML decodable codes of similar decoding complexity. These results of this paper have brought to light the following problems that need to be addressed.

- The rates of the DSTBCs of this paper are upper bounded by 1 cspcu since we have considered relay networks with only a single antenna source. To achieve higher rates in the network, one has to employ a multiple antenna source. How does one design full-diversity PIC/PIC-SIC decodable DSTBCs in this case?
- Does the full diversity criterion given in this paper hold for PIC/PIC-SIC decoding in multihop wireless networks also? How does the criterion vary in non-coherent and/or asynchronous networks?

- It is known that the power allocation $\pi_1 = 1$ and $\pi_2 = \frac{1}{R}$ is optimal for ML decoding at the destination [2]. Is the same true for PIC/PIC-SIC decoding as well?

ACKNOWLEDGMENT

This work was supported partly by the DRDO-IISc program on Advanced Research in Mathematical Engineering through a research grant, and partly by the INAE Chair Professorship grant to B. S. Rajan.

## APPENDIX A

### PROOF OF THEOREM 1

Before giving the proof of Theorem 1, we present a few results to be used in the proof.

Let $\Gamma'$ be the covariance matrix of the noise vector $vec(U)$ given in (1).

*Proposition 3:* Let $\Gamma' = \begin{bmatrix} C_{1,1} & C_{1,2} & \cdots & C_{1,N_D} \\ C_{2,1} & C_{2,2} & \cdots & C_{2,N_D} \\ \vdots & \vdots & \ddots & \vdots \\ C_{N_D,1} & C_{N_D,2} & \cdots & C_{N_D,N_D} \end{bmatrix}$, where the submatrices $C_{l_1,l_2} \in \mathbb{C}^{T_2 \times T_2}$, $1 \leq l_1, l_2 \leq N_D$. Then $C_{l_1,l_2} = \frac{\pi_2 P}{\pi_1 P + 1} \sum_{j=1}^{N} g_{j,l_1} g_{j,l_2}^* \bar{B}_j \bar{B}_j^H + \mathbf{1}\{l_1 = l_2\} I_{T_2}$, for $1 \leq l_1, l_2 \leq N_D$ where $\mathbf{1}\{\cdot\}$ is the indicator function.

*Proof:* For $1 \leq l_1, l_2 \leq N_D$, $C_{l_1,l_2} = \mathsf{E}\left(u_{l_1} u_{l_2}^H\right)$. Expanding with the help of (1) and using the fact that the Gaussian vectors $v_j$, $j = 1, \ldots, N$, and $w_l$, $l = 1, \ldots, N_D$ are of zero mean and are mutually independent we have, $C_{l_1,l_2} = \mathsf{E}\left\{\sum_{j_1=1}^{N} \sum_{j_2=1}^{N} \frac{\pi_2 P}{\pi_1 P + 1} g_{j_1,l_1} g_{j_2,l_2}^* \bar{B}_{j_1} \bar{v}_{j_1} \bar{v}_{j_2}^H \bar{B}_{j_2}^H + w_{l_1} w_{l_2}^H\right\}$
$= \sum_{j_1=1}^{N} \sum_{j_2=1}^{N} \frac{\pi_2 P}{\pi_1 P + 1} g_{j_1,l_1} g_{j_2,l_2}^* \bar{B}_{j_1} \mathsf{E}\left(\bar{v}_{j_1} \bar{v}_{j_2}^H\right) \bar{B}_{j_2}^H + \mathsf{E}\left(w_{l_1} w_{l_2}^H\right)$. Using the fact that $\mathsf{E}\left(\bar{v}_{j_1} \bar{v}_{j_2}^H\right) = \mathbf{1}\{j_1 = j_2\} I_{T_1}$ and $\mathsf{E}\left(w_{l_1} w_{l_2}^H\right) = \mathbf{1}\{l_1 = l_2\} I_{T_2}$, we get the desired result. ∎

For any two square matrices $A$ and $B$ of same dimension, the notation $A \preceq B$ implies that $B - A$ is a positive semidefinite (PSD) matrix. Let $\beta = max\left\{||\bar{B}_j||_F^2 | \ j = 1, \ldots, N\right\}$ and $\alpha = T_2 N_D + \frac{\beta \pi_2 P}{\pi_1 P + 1} \sum_{j=1}^{N} \sum_{l=1}^{N_D} |g_{j,l}|^2$. We use the following proposition to prove Theorem 1.

*Proposition 4:* The covariance matrix of $\widetilde{vec}(U)$, $\Gamma$ satisfies $\Gamma \preceq \alpha I_{2T_2 N_D}$.

*Proof:* Since $\Gamma$ is a covariance matrix, it is PSD and hence has a complete set of eigenvalues. If $\lambda_i$, $i = 1, \ldots, 2T_2 N_D$, are the eigenvalues of $\Gamma$, we have: $\Gamma \preceq \sum_{i=1}^{2T_2 N_D} \lambda_i I = Tr(\Gamma) I$, where $Tr(\cdot)$ represents the trace of a matrix. Since $\Gamma = \frac{1}{2}\begin{bmatrix} \Gamma'_{Re} & -\Gamma'_{Im} \\ \Gamma'_{Im} & \Gamma'_{Re} \end{bmatrix}$, we have $Tr(\Gamma) = Tr(\Gamma'_{Re})$. Note that $\Gamma'$ is itself a covariance matrix and hence is Hermitian. Thus, all of its diagonal entries are real and hence, $Tr(\Gamma'_{Re}) = Tr(\Gamma')$. From Proposition 3, we have: $Tr(\Gamma) = \sum_{l=1}^{N_D} Tr(C_{l,l})$. Thus,

$$Tr(\Gamma) = \sum_{l=1}^{N_D} \left[\frac{\pi_2 P}{\pi_1 P + 1} \sum_{j=1}^{N} |g_{j,l}|^2 Tr(\bar{B}_j \bar{B}_j^H) + T_2\right]$$
$$= \frac{\pi_2 P}{\pi_1 P + 1} \sum_{j=1}^{N} \sum_{l=1}^{N_D} |g_{j,l}|^2 ||\bar{B}_j||_F^2 + T_2 N_D$$
$$\leq \frac{\pi_2 P}{\pi_1 P + 1} \sum_{j=1}^{N} \sum_{l=1}^{N_D} |g_{j,l}|^2 \beta + T_2 N_D = \alpha.$$

Thus, $\Gamma \preceq Tr(\Gamma) I_{2T_2 N_D} \preceq \alpha I_{2T_2 N_D}$. ∎

Let $m$ be a positive integer, $V'$ be a subspace of $\mathbb{R}^m$ and let $A \in \mathbb{R}^{m \times m}$ be any symmetric full-rank matrix. We have the following result.

*Proposition 5:* If $V = AV' = \{Av | v \in V'\}$, then $V^\perp = A^{-1}V'^\perp$.

*Proof:* We have, $V^\perp = (AV')^\perp = \{w | w^T A v = 0 \ \forall v \in V'\} = \{w | (A^T w)^T v = 0 \ \forall v \in V'\}$. Replacing $A^T w = Aw$ by $u$, we have, $V^\perp = \{A^{-1} u | u^T v = 0 \ \forall v \in V'\} = A^{-1}\{u | u^T v = 0 \ \forall v \in V'\} = A^{-1} V'^\perp$. ∎

*Proof of Theorem 1:* Let

$$\mathcal{C} = \left\{\sum_{i=1}^{K} x_i A_i \Big| x_{\mathcal{I}_k} \in \mathcal{A}_{\mathcal{I}_k}, k = 1, \ldots, g\right\},$$

and for a given $N \times N_D$ complex matrix $H$ and positive real number $\rho$, let $G' = \sqrt{\rho}[\widetilde{vec}(A_1 H) \cdots \widetilde{vec}(A_K H)] = [g'_1 \, g'_2 \cdots g'_K]$. For $k = 1, \ldots, g$, let $V'_{\mathcal{I}_k}$ be the column space of the matrix $G'_{\mathcal{I}_k^c}$ and $P'_{\mathcal{I}_k}$ be the matrix that projects a vector onto the subspace $V'^{\perp}_{\mathcal{I}_k}$.

*Theorem 3 ([8]):* If for every $X_1, X_2 \in \mathcal{C}$ and $X_1 \neq X_2$, $rank(X_1 - X_2)$ is $N$ and for every $k = 1, \ldots, g$, every $H \neq \mathbf{0}$ and every $a_k \in \Delta \mathcal{A}_{\mathcal{I}_k} \setminus \{0\}$, $G'_{\mathcal{I}_k} a_k \notin V'_{\mathcal{I}_k}$, then, there exists a real number $c > 0$ such that, for any $k = 1, \ldots, g$, any $a_k \in \Delta \mathcal{A}_{\mathcal{I}_k} \setminus \{0\}$ and $H \neq \mathbf{0}$, we have: $||P'_{\mathcal{I}_k} G'_{\mathcal{I}_k} a_k||_F^2 > c\rho ||H||_F^2$.

The result of Theorem 3 is independent of the statistics of $H$. The matrix $G'$ in Theorem 3 is identical to the matrix (3), which arises during the PIC decoding of the DSTBC $\mathcal{C}$ in a relay network with $N$ relays and $N_D$ antennas at the destination when $\rho = \frac{\pi_1 \pi_2 P^2}{\pi_1 P + 1}$. Hence, the result of Theorem 3 can be used to prove diversity results for the relay network if the two conditions of the theorem are satisfied. Now, the criterion in the hypothesis of Theorem 1 is same as the criterion given in [9] for the code $\mathcal{C}$ to achieve full-diversity in the point-to-point MIMO channel as an STBC with PIC decoding under the grouping scheme $\mathcal{I}_1, \ldots, \mathcal{I}_g$. Further, it is shown in [9] that this criterion is equivalent to the sufficient condition given in Theorem 3. Thus, with $\rho = \frac{\pi_1 \pi_2 P^2}{\pi_1 P + 1}$, for any $k = 1, \ldots, g$, any $a_k \in \Delta \mathcal{A}_{\mathcal{I}_k} \setminus \{0\}$ and any channel realization $H \neq \mathbf{0}$ of the relay network, we have:

$$||P'_{\mathcal{I}_k} G'_{\mathcal{I}_k} a_k||_F^2 > c\rho ||H||_F^2. \quad (7)$$

Let $k \in \{1, \ldots, g\}$. We are interested in deriving an upper bound on the pairwise error probability during PIC decoding (4) of the $k^{th}$ symbol vector $x_{\mathcal{I}_k}$. The received vector $y$ satisfies $y = Gx + n = \sum_{k=1}^{g} G_{\mathcal{I}_k} x_{\mathcal{I}_k} + n$. Since $P_{\mathcal{I}_k}$ is the projection onto the subspace which is orthogonal to the subspace spanned by the column vectors of $G_{\mathcal{I}_\ell}$, $1 \leq \ell \leq g$ and $\ell \neq k$, we have that $P_{\mathcal{I}_k} G_{\mathcal{I}_\ell} = \mathbf{0}$ for $1 \leq \ell \leq g, \ell \neq k$. Thus, $P_{\mathcal{I}_k} y = \sum_{\ell=1}^{g} P_{\mathcal{I}_k} G_{\mathcal{I}_\ell} x_{\mathcal{I}_\ell} + P_{\mathcal{I}_k} n = P_{\mathcal{I}_k} G_{\mathcal{I}_k} x_{\mathcal{I}_k} + P_{\mathcal{I}_k} n$. The PIC decoder (4) is:

$arg \ min_{\tilde{x}_{\mathcal{I}_k} \in \mathcal{A}_{\mathcal{I}_k}} ||P_{\mathcal{I}_k} y P_{\mathcal{I}_k} G_{\mathcal{I}_k} \tilde{x}_{\mathcal{I}_k}||_F^2$
$= arg \ min_{\tilde{x}_{\mathcal{I}_k} \in \mathcal{A}_{\mathcal{I}_k}} ||P_{\mathcal{I}_k} G_{\mathcal{I}_k} x_{\mathcal{I}_k} + P_{\mathcal{I}_k} n - P_{\mathcal{I}_k} G_{\mathcal{I}_k} \tilde{x}_{\mathcal{I}_k}||_F^2$
$= arg \ min_{\tilde{x}_{\mathcal{I}_k} \in \mathcal{A}_{\mathcal{I}_k}} ||P_{\mathcal{I}_k} G_{\mathcal{I}_k} (x_{\mathcal{I}_k} - \tilde{x}_{\mathcal{I}_k}) + P_{\mathcal{I}_k} n||_F^2.$

The Gaussian noise vector $n$ is white and has zero mean. Since $P_{\mathcal{I}_k}$ is the projection onto $V^{\perp}_{\mathcal{I}_k}$, the noise vector $P_{\mathcal{I}_k} n$ has no components in the subspace $V_{\mathcal{I}_k}$ and the component of $P_{\mathcal{I}_k} n$ along the subspace $V^{\perp}_{\mathcal{I}_k}$ is a white Gaussian noise with zero mean and unit covariance. The difference signal vector $P_{\mathcal{I}_k} G_{\mathcal{I}_k} (x_{\mathcal{I}_k} - \tilde{x}_{\mathcal{I}_k})$ also lies in the subspace $V^{\perp}_{\mathcal{I}_k}$. Thus, the probability that the PIC decoder will decide in favor of $\tilde{x}_{\mathcal{I}_k}$ when the symbol $x_{\mathcal{I}_k}$ is transmitted, given the channel realization $H$, is $\mathsf{PEP}(x_{\mathcal{I}_k} \to \tilde{x}_{\mathcal{I}_k} | H) = \mathcal{Q}\left(\frac{||P_{\mathcal{I}_k} G_{\mathcal{I}_k} a_k||_F}{2}\right)$, where $\mathcal{Q}(\cdot)$ is the Gaussian tail function and $a_k = x_{\mathcal{I}_k} - \tilde{x}_{\mathcal{I}_k}$. Using the Chernoff bound on the $\mathcal{Q}$ function, we have

$$\mathsf{PEP}(x_{\mathcal{I}_k} \to \tilde{x}_{\mathcal{I}_k} | H) \leq exp\left(\frac{-||P_{\mathcal{I}_k} G_{\mathcal{I}_k} a_k||_F^2}{4}\right). \quad (8)$$

In order to derive a lower bound on $||P_{\mathcal{I}_k} G_{\mathcal{I}_k} a_k||_F^2$, we now express $P_{\mathcal{I}_k}$ and $G_{\mathcal{I}_k}$ in terms of $P'_{\mathcal{I}_k}$ and $G'_{\mathcal{I}_k}$. Let $A = \Gamma^{-\frac{1}{2}}$ denote the square root of $\Gamma^{-1}$. Since $\Gamma$ and $\Gamma^{-1}$ are PSD and symmetric, $A$ can be chosen to be the unique PSD symmetric square root of $\Gamma^{-1}$ [17]. From Proposition 4, we have $A \succeq \frac{1}{\sqrt{\alpha}} I_{2T_2 N_D}$. Let $Q^T_{\mathcal{I}_k}$ be a matrix whose columns form an orthonormal basis of $V^{\perp}_{\mathcal{I}_k}$ and $Q'^T_{\mathcal{I}_k}$ be a matrix whose columns form an orthonormal basis of $V'^{\perp}_{\mathcal{I}_k}$. Thus, $P_{\mathcal{I}_k} = Q^T_{\mathcal{I}_k} Q_{\mathcal{I}_k}$ and $P'_{\mathcal{I}_k} = Q'^T_{\mathcal{I}_k} Q'_{\mathcal{I}_k}$. Also, for any vector $v \in \mathbb{R}^{2T_2 N_D}$, we have $||P_{\mathcal{I}_k} v||_F = ||Q_{\mathcal{I}_k} v||_F$ and $||P'_{\mathcal{I}_k} v||_F = ||Q'_{\mathcal{I}_k} v||_F$. Since $G = AG'$, it is clear that $G_{\mathcal{I}_k} = AG'_{\mathcal{I}_k}$ and $V_{\mathcal{I}_k} = AV'_{\mathcal{I}_k} = \{Av | v \in V'_{\mathcal{I}_k}\}$. From Proposition 5, it is clear that $V^{\perp}_{\mathcal{I}_k}$ is spanned by the column vectors of the matrix $A^{-1} Q'^T_{\mathcal{I}_k}$. Thus, $P_{\mathcal{I}_k} = A^{-1} Q'^T_{\mathcal{I}_k} \left(Q'_{\mathcal{I}_k} A^{-1^T} A^{-1} Q'^T_{\mathcal{I}_k}\right)^{-1} Q'_{\mathcal{I}_k} A^{-1^T} = A^{-1} Q'^T_{\mathcal{I}_k} \left(Q'_{\mathcal{I}_k} A^{-2} Q'^T_{\mathcal{I}_k}\right)^{-1} Q'_{\mathcal{I}_k} A^{-1}$.

Let $k \in \{1, \ldots, g\}$ and $a_k \in \Delta \mathcal{A}_{\mathcal{I}_k} \setminus \{0\}$. Consider $||P_{\mathcal{I}_k} G_{\mathcal{I}_k} a_k||_F^2 = a_k^T G^T_{\mathcal{I}_k} P^T_{\mathcal{I}_k} P_{\mathcal{I}_k} G_{\mathcal{I}_k} a_k$. Using $G_{\mathcal{I}_k} = AG'_{\mathcal{I}_k}$ and expanding $P_{\mathcal{I}_k}$ as above, we get

$$||P_{\mathcal{I}_k} G_{\mathcal{I}_k} a_k||_F^2 = a_k^T G'^T_{\mathcal{I}_k} Q'^T_{\mathcal{I}_k} \left(Q'_{\mathcal{I}_k} A^{-2} Q'^T_{\mathcal{I}_k}\right)^{-1} Q'_{\mathcal{I}_k} G'_{\mathcal{I}_k} a_k$$
$$= ||\left(Q'_{\mathcal{I}_k} A^{-2} Q'^T_{\mathcal{I}_k}\right)^{-\frac{1}{2}} Q'_{\mathcal{I}_k} G'_{\mathcal{I}_k} a_k||_F^2.$$

Since $A \succeq \frac{1}{\sqrt{\alpha}} I_{2T_2 N_D}$ and the rows of $Q'_{\mathcal{I}_k}$ are orthonormal, we have $\left(Q'_{\mathcal{I}_k} A^{-2} Q'^T_{\mathcal{I}_k}\right)^{-\frac{1}{2}} \succeq \frac{1}{\sqrt{\alpha}} (Q'_{\mathcal{I}_k} Q'^T_{\mathcal{I}_k})^{-\frac{1}{2}} = \frac{1}{\sqrt{\alpha}} I$. Thus,

$$||P_{\mathcal{I}_k} G_{\mathcal{I}_k} a_k||_F^2 \geq ||\frac{1}{\sqrt{\alpha}} I Q'_{\mathcal{I}_k} G'_{\mathcal{I}_k} a_k||_F^2 = \frac{1}{\alpha} ||Q'_{\mathcal{I}_k} G'_{\mathcal{I}_k} a_k||_F^2$$
$$= \frac{1}{\alpha} ||P'_{\mathcal{I}_k} G'_{\mathcal{I}_k} a_k||_F^2 > \frac{c\rho ||H||_F^2}{\alpha}.$$

The last step follows from (7). Thus, for any $k = 1, \ldots, g$ and $a_k \in \Delta \mathcal{A}_{\mathcal{I}_k} \setminus \{0\}$, we have that $||P_{\mathcal{I}_k} G_{\mathcal{I}_k} a_k||_F^2 > \frac{c\rho ||H||_F^2}{\alpha}$. Using this inequality with (8) we get $\mathsf{PEP}(x_{\mathcal{I}_k} \to \tilde{x}_{\mathcal{I}_k} | H)$

$$\leq exp\left(-\frac{c\rho ||H||_F^2}{4\alpha}\right).$$

Note that $||H||_F^2 = ||F\mathcal{G}||_F^2 = \sum_{j=1}^{N} |f_j|^2 \left(\sum_{l=1}^{N_D} |g_{j,l}|^2\right)$. The squared absolute values of the channel gains $|f_j|^2$ and $|g_{j,l}|^2$ are all independent of each other and are exponential random variables with unit mean. Let $t_j = \sum_{l=1}^{N_D} |g_{j,l}|^2$, for $j = 1, \ldots, N$. Then, the random variables $|f_j|^2 t_j$, $j = 1, \ldots, N_D$ are independent of each other. Further, $exp\left(-\frac{c\rho ||H||_F^2}{4\alpha}\right) = \prod_{j=1}^{N} exp\left(-\frac{c\rho |f_j|^2 t_j}{4\alpha}\right)$. Since $|f_j|^2$ is exponentially distributed with unit mean, for any $s > 0$, we have $\mathsf{E}(exp(-s|f_j|^2)) = \frac{1}{1+s}$, for $j = 1, \ldots, N$. Thus, the average pairwise error probability,

$$\text{PEP}(x_{\mathcal{I}_k} \to \tilde{x}_{\mathcal{I}_k}) \leq \mathsf{E}\left(\prod_{j=1}^{N} exp\left(-\frac{c\rho|f_j|^2 t_j}{4\alpha}\right)\right)$$
$$= \mathsf{E}\left(\prod_{j=1}^{N}\left(1 + \frac{c\rho t_j}{4\alpha}\right)^{-1}\right).$$

Substituting the values for $\rho$ and $\alpha$ as $\rho = \frac{\pi_1 \pi_2 P^2}{\pi_1 P + 1}$ and $\alpha = T_2 N_D + \frac{\beta \pi_2 P}{\pi_1 P + 1} \sum_{j'=1}^{N} t_{j'}$, we get

$$1 + \frac{c\rho t_j}{4\alpha} = 1 + \frac{c\pi_1 \pi_2 P^2 t_j}{4\left[(\pi_1 P + 1)T_2 N_D + \beta \pi_2 P \sum_{j'=1}^{N} t_{j'}\right]}.$$

For large $P$, $\pi_1 P + 1 \approx \pi_1 P$. Using this approximation and on further simplification, we get $\text{PEP}(x_{\mathcal{I}_k} \to \tilde{x}_{\mathcal{I}_k})$

$$\leq \mathsf{E}\left(\prod_{j=1}^{N}\left[1 + \frac{c\pi_2 P}{4 T_2 N_D} \frac{t_j}{1 + \frac{\beta \pi_2}{\pi_2 T_2 N_D} \sum_{j'=1}^{N} t_{j'}}\right]^{-1}\right). \quad (9)$$

In Theorem 4 of [2] an upper bound for an expression of which (9) is a special case is given. The result in [2] for the special case (9) is as follows.

*Theorem 4 ([2]):* The pairwise error probability (9) can be upper bounded by $c_0 P^{-d}$, where $c_0$ is a positive real number and $d = N\left(1 - \frac{log(logP)}{logP}\right)$ if $N_D = 1$ and $d = N$ if $N_D > 1$.

From Theorem 4, it is clear that the DSTBC $\mathcal{C}$ achieves a diversity of $d$ with PIC decoding, where $d = N\left(1 - \frac{log(logP)}{logP}\right)$ if $N_D = 1$ and $d = N$ if $N_D > 1$. This completes the proof of Theorem 1. ∎

## APPENDIX B
### OUTLINE OF THE PROOF OF THEOREM 2

Let $d = \mathbf{1}\{N_D = 1\} \cdot N\left(1 - \frac{log(logP)}{logP}\right) + \mathbf{1}\{N_D > 1\} \cdot N$ and let $\mathsf{P}(\cdot)$ denote the probability of an event. For $k = 1, \ldots, g$, let $E_k$ denote the event that the $k^{th}$ information symbol vector $x_{\mathcal{I}_k}$ is erroneously decoded by the PIC-SIC decoder. We want to prove that $\mathsf{P}(E_1 \cup \cdots \cup E_g) \leq c_0 P^{-d}$, for large $P$ and for some positive real number $c_0$. For $k = 1, \ldots, g$, $\mathsf{P}(E_k)$ satisfies (10) given at the top of the next page. It is enough to show that $\mathsf{P}(E_k|E_1^c \cap \cdots \cap E_{k-1}^c) \leq c_k P^{-d}$ for $k = 1, \ldots, g$ and some set of positive real numbers $\{c_k\}$. Then, from (10), it can be shown using recursion that
$\mathsf{P}(E_1 \cup \cdots \cup E_g) \leq \sum_{k=1}^{g} \mathsf{P}(E_k) \leq c_0 P^{-d}$, for some $c_0 > 0$. We now derive an upper bound for $\mathsf{P}(E_k|E_1^c \cap \cdots \cap E_{k-1}^c)$, the probability of erroneously decoding the $k^{th}$ symbol vector when all the previous symbol vectors have been decoded correctly. For a given $N \times N_D$ complex matrix $H$ and real number $\rho$, let $G' = \sqrt{\rho}[\widetilde{vec}(A_1 H) \cdots \widetilde{vec}(A_K H)] = [g_1' \ g_2' \cdots g_K']$. For $k = 1, \ldots, g$, let $\tilde{V}'_{\mathcal{I}_k}$ be the column space of the matrix $G'_{\mathcal{I}_k}$ and $\tilde{P}'_{\tilde{\mathcal{I}}_k}$ be the matrix that projects a vector onto the subspace $\tilde{V}'^{\perp}_{\mathcal{I}_k}$.

*Theorem 5 ([8]):* If for any $X_1, X_2 \in \mathcal{C}$ and $X_1 \neq X_2$, $rank(X_1 - X_2)$ is $N$ and for every $k = 1, \ldots, g$, every $H \neq \mathbf{0}$ and every $a_k \in \Delta \mathcal{A}_{\mathcal{I}_k} \setminus \{0\}$, $G'_{\tilde{\mathcal{I}}_k} a_k \notin \tilde{V}'_{\mathcal{I}_k}$, then, there exists a real number $c > 0$ such that, for any $k = 1, \ldots, g$, any $a_k \in \Delta \mathcal{A}_{\mathcal{I}_k} \setminus \{0\}$ and $H \neq \mathbf{0}$, we have: $||\tilde{P}'_{\mathcal{I}_k} G'_{\tilde{\mathcal{I}}_k} a_k||_F^2 > c\rho ||H||_F^2$.

The criterion in the hypothesis of Theorem 2 is same as the criterion given in [9] for the STBC $\mathcal{C}$ to achieve full-diversity in the point-to-point MIMO channel with PIC-SIC decoding under the grouping scheme $\mathcal{I}_1, \ldots, \mathcal{I}_g$. Further, it is shown in [9] that this criterion is equivalent to the sufficient condition given in Theorem 5. Thus, for any $k = 1, \ldots, g$, any $a_k \in \Delta \mathcal{A}_{\mathcal{I}_k} \setminus \{0\}$ and any channel realization $H \neq \mathbf{0}$ of the relay network, we have $||\tilde{P}'_{\mathcal{I}_k} G'_{\tilde{\mathcal{I}}_k} a_k||_F^2 > c\rho ||H||_F^2$, where $\rho = \frac{\pi_1 \pi_2 P^2}{\pi_1 P + 1}$.

Consider the $k^{th}$ iteration in the PIC-SIC decoding algorithm given in Section III-A in the case where all the previous symbol vectors $x_{\mathcal{I}_1}, \ldots, x_{\mathcal{I}_{k-1}}$ have been decoded correctly. Due to successive interference cancellation (Step 2 of the PIC-SIC decoding algorithm in Section III-A), the signal $y_k$ will only have a noisy version of the contributions from symbol vectors $x_{\mathcal{I}_k}, \ldots, x_{\mathcal{I}_g}$, i.e., $y_k = \sum_{\ell=k}^{g} G_{\mathcal{I}_\ell} x_{\mathcal{I}_\ell} + n$. Since $\tilde{P}_{\mathcal{I}_k}$ is the projection onto the subspace which is orthogonal to the subspace spanned by the column vectors of $G_{\mathcal{I}_\ell}$, $k < \ell \leq g$, we have that $\tilde{P}_{\mathcal{I}_k} G_{\mathcal{I}_\ell} = \mathbf{0}$ for $k < \ell \leq g$. Thus, $\tilde{P}_{\mathcal{I}_k} y_k = \sum_{\ell=k}^{g} \tilde{P}_{\mathcal{I}_k} G_{\mathcal{I}_\ell} x_{\mathcal{I}_\ell} + \tilde{P}_{\mathcal{I}_k} n = \tilde{P}_{\mathcal{I}_k} G_{\mathcal{I}_k} x_{\mathcal{I}_k} + \tilde{P}_{\mathcal{I}_k} n$. Hence, the output of PIC-SIC decoder in the $k^{th}$ iteration is

$$arg \ min_{\tilde{x}_{\mathcal{I}_k} \in \mathcal{A}_{\mathcal{I}_k}} ||\tilde{P}_{\mathcal{I}_k} y_k - \tilde{P}_{\mathcal{I}_k} G_{\mathcal{I}_k} \tilde{x}_{\mathcal{I}_k}||_F^2$$
$$= arg \ min_{\tilde{x}_{\mathcal{I}_k} \in \mathcal{A}_{\mathcal{I}_k}} ||\tilde{P}_{\mathcal{I}_k} G_{\mathcal{I}_k} x_{\mathcal{I}_k} + \tilde{P}_{\mathcal{I}_k} n - \tilde{P}_{\mathcal{I}_k} G_{\mathcal{I}_k} \tilde{x}_{\mathcal{I}_k}||_F^2$$
$$= arg \ min_{\tilde{x}_{\mathcal{I}_k} \in \mathcal{A}_{\mathcal{I}_k}} ||\tilde{P}_{\mathcal{I}_k} G_{\mathcal{I}_k} (x_{\mathcal{I}_k} - \tilde{x}_{\mathcal{I}_k}) + \tilde{P}_{\mathcal{I}_k} n||_F^2.$$

Using an argument similar to the proof of Theorem 1, the probability that the PIC-SIC decoder will decide in favor of $\tilde{x}_{\mathcal{I}_k}$ when the symbol $x_{\mathcal{I}_k}$ is transmitted, given the channel realization $H$, can be shown to be $\text{PEP}(x_{\mathcal{I}_k} \to \tilde{x}_{\mathcal{I}_k}|H, E_1^c \cap \cdots \cap E_{k-1}^c) = \mathcal{Q}\left(\frac{||\tilde{P}_{\mathcal{I}_k} G_{\mathcal{I}_k} a_k||_F}{2}\right)$, where $a_k = x_{\mathcal{I}_k} - \tilde{x}_{\mathcal{I}_k}$. Using the Chernoff bound on the $\mathcal{Q}$ function, we have

$$\text{PEP}(x_{\mathcal{I}_k} \to \tilde{x}_{\mathcal{I}_k}|H, E_1^c \cap \cdots \cap E_{k-1}^c) \leq exp\left(\frac{-||\tilde{P}_{\mathcal{I}_k} G_{\mathcal{I}_k} a_k||_F^2}{4}\right).$$

Using an argument similar to the proof of Theorem 1, it can be shown that

$$||\tilde{P}_{\mathcal{I}_k} G_{\mathcal{I}_k} a_k||_F^2 \geq \frac{1}{\alpha} ||\tilde{P}'_{\mathcal{I}_k} G'_{\tilde{\mathcal{I}}_k} a_k||_F^2 > \frac{c\rho ||H||_F^2}{\alpha},$$

and then the average pairwise error probability, $\text{PEP}(x_{\mathcal{I}_k} \to \tilde{x}_{\mathcal{I}_k}|E_1^c \cap \cdots \cap E_{k-1}^c)$ can be shown to be upper bounded by $b_k P^{-d}$ for some $b_k > 0$. This completes the proof. ∎

## APPENDIX C
### RESISTANCE TO RELAY NODE FAILURES

Consider the DSTBC $\mathcal{C}$ in (2), designed for a relay network with $N$ relays and which satisfies the full-diversity condition in Theorem 1 or Theorem 2. For some $a \in \{1, \ldots, N\}$, suppose $a$ number of relay nodes stop participating in the cooperative

$$
\begin{aligned}
\mathsf{P}(E_k) &= \mathsf{P}(E_k|E_1^c\cap\cdots\cap E_{k-1}^c)\mathsf{P}(E_1^c\cap\cdots\cap E_{k-1}^c) + \mathsf{P}(E_k|E_1\cup\cdots\cup E_{k-1})\mathsf{P}(E_1\cup\cdots\cup E_{k-1}) \\
&\le \mathsf{P}(E_k|E_1^c\cap\cdots\cap E_{k-1}^c)\cdot 1 + 1\cdot \mathsf{P}(E_1\cup\cdots\cup E_{k-1}) \\
&\le \mathsf{P}(E_k|E_1^c\cap\cdots\cap E_{k-1}^c) + \sum_{k'=1}^{k-1}\mathsf{P}(E_{k'}).
\end{aligned} \quad (10)
$$

protocol. This may happen when the nodes move out of the network or are switched off. Also, let the destination be aware of the nodes that are currently participating in the cooperative transmission. Then, the DSTBC $\mathcal{C}'$ seen by the destination is the code $\mathcal{C}$ with the $a$ columns corresponding to the failed relay nodes dropped from each codeword matrix. One would like the new DSTBC $\mathcal{C}'$ to provide full-diversity in the modified relay network with $N-a$ relays. This ensures that good error performance is maintained in the network with minimum protocol overhead when a subset of relay nodes stop participating.

*Proposition 6:* Let $\mathcal{C}$ satisfy the full-diversity criterion of Theorem 1 (Theorem 2) and let the destination be employed with a PIC (PIC-SIC) decoder. Then the DSTBC $\mathcal{C}'$ provides full-diversity with PIC (PIC-SIC) decoding for the modified network with $N-a$ relays.

*Proof:* We give the proof for the case when the destination employs a PIC decoder. The proof for PIC-SIC decoder is similar. For $i=1,\ldots,K$, let $A'_i$ be the $T_2\times(N-a)$ matrix formed by dropping the $a$ columns corresponding to the failed relay nodes from the matrix $A_i$. Let the grouping scheme be $\mathcal{I}_1,\ldots,\mathcal{I}_g$. The DSTBC for the modified network satisfies $\mathcal{C}'=\left\{\sum_{i=1}^{K}x_iA'_i|x_{\mathcal{I}_k}\in\mathcal{A}_{\mathcal{I}_k},\ k=1,\ldots,g\right\}$, which is obtained from the design $\mathbf{X}'=\sum_{i=1}^{K}x_iA'_i$. Also, for every $k=1,\ldots,g$, the rank of $X_{\mathcal{I}_k}(a_k)+X_{\mathcal{I}_k^c}(u)$ is $N$ for every $a_k\in\Delta\mathcal{A}_{\mathcal{I}_k}\setminus\{0\}$ and every $u\in\mathbb{R}^{|\mathcal{I}_k^c|}$. The $N$ columns of the matrix $X_{\mathcal{I}_k}(a_k)+X_{\mathcal{I}_k^c}(u)$ are linearly independent over $\mathbb{C}$. Thus, the $N-a$ columns of the matrix $X'_{\mathcal{I}_k}(a_k)+X'_{\mathcal{I}_k^c}(u)$, formed by dropping $a$ columns from $X_{\mathcal{I}_k}(a_k)+X_{\mathcal{I}_k^c}(u)$, are also linearly independent. Hence, the rank of $X'_{\mathcal{I}_k}(a_k)+X'_{\mathcal{I}_k^c}(u)$ is $N-a$ for every $k=1,\ldots,g$, $a_k\in\Delta\mathcal{A}_{\mathcal{I}_k}\setminus\{0\}$ and every $u\in\mathbb{R}^{|\mathcal{I}_k^c|}$. Thus, from Theorem 1, the DSTBC $\mathcal{C}'$ achieves full-diversity in the modified network with $N-a$ relays and PIC decoding. ∎

Proposition 6 also tells us that new full-diversity DSTBCs for relay networks with $N-a$ relays can be obtained by simply dropping any set of $a$ columns from a known full-diversity DSTBC for a network with $N$ relays.

## APPENDIX D
## PROOF OF PROPOSITION 1

When $\lambda=1$ in the proposed design (6), for each $p=1,\ldots,n$ we have $\mathbf{W}(p,1)=\mathbf{W}(p,2)=\cdots=\mathbf{W}(p,L)$. Thus, the first diagonal layer encodes the first $K'$ symbols $x_1,\ldots,x_{K'}$, the second diagonal layer encodes the second $K'$ symbols $x_{K'+1},\ldots,x_{2K'}$ and so on. Now consider any $u=[u_1,\ldots,u_K]\in\mathbb{R}^K\setminus\{0\}$. We need to show that $X=\sum_{i=1}^{K}u_iA_i$ is of full-rank. Let the weight matrices of $\mathbf{W}$ be $A'_1,\ldots,A'_{K'}$. From (6), it is clear that

$$
X = \begin{bmatrix}
W(1,1) & 0 & \cdots & 0 \\
W(2,1) & W(1,1) & \cdots & \vdots \\
\vdots & W(2,1) & \ddots & 0 \\
\vdots & \vdots & \ddots & W(1,1) \\
\vdots & \vdots & \cdots & W(2,1) \\
\vdots & \vdots & \cdots & \vdots \\
W(n,1) & \vdots & \cdots & \vdots \\
0 & W(n,1) & \cdots & \vdots \\
\vdots & \vdots & \ddots & \vdots \\
0 & 0 & \cdots & W(n,1)
\end{bmatrix},
$$

wherein $W(p,1)=\sum_{j=1}^{K'}u_{K'(p-1)+j}A'_j$ for $p=1,\ldots,n$.

Let $k$ be the smallest integer such that $u_k\ne 0$ and let the $k^{th}$ symbol $x_k$ be encoded by the $m^{th}$ layer, i.e., $K'(m-1)+1\le k\le K'm$. Thus, $W(1,1)=W(2,1)=\cdots=W(m-1,1)=\mathbf{0}$. Then $X$ equals the matrix

$$
\begin{bmatrix}
0 & 0 & \cdots & 0 \\
\vdots & \vdots & \vdots & \vdots \\
0 & 0 & \cdots & 0 \\
W(m,1) & 0 & \cdots & 0 \\
W(m+1,1) & W(m,1) & \cdots & \vdots \\
\vdots & W(m+1,1) & \ddots & 0 \\
\vdots & \vdots & \ddots & W(m,1) \\
\vdots & \vdots & \cdots & W(m+1,1) \\
\vdots & \vdots & \cdots & \vdots \\
W(n,1) & \vdots & \cdots & \vdots \\
0 & W(n,1) & \cdots & \vdots \\
\vdots & \vdots & \ddots & \vdots \\
0 & 0 & \cdots & W(n,1)
\end{bmatrix}. \quad (11)
$$

The first $(m-1)$ diagonal layers of (11) have only all-zero matrices. Since $\mathbf{W}$ is a COD its weight matrices satisfy $A'^H_iA'_j+A'^H_jA'_i=2\delta_{i,j}I_{N'}$ [15]. Using this property of CODs and the fact that $k\in\{K'(m-1)+1,K'(m-1)+2,\ldots,K'm\}$,

we have

$$det\left(W(m,1)^H W(m,1)\right) = det\left(\sum_{j=1}^{K'} u_{K'(m-1)+j}^2 I_{N'}\right)$$

$$= \left(\sum_{j=1}^{K'} u_{K'(m-1)+j}^2\right)^{N'} \geq u_k^{2N'} > 0, \quad (12)$$

where $det(\cdot)$ represents the determinant of a matrix. Thus, $W(m,1)$ is full-ranked. Using (11) and the fact that $W(m,1)$ is full-ranked, it is straightforward to show that $X$ is full-ranked. From Corollary 1, the DSTBC achieves full diversity with ZF and ZF-SIC decoding. ∎

## APPENDIX E
### PROOF OF PROPOSITION 2

We prove the part of the claim concerning PIC-SIC decoding. The proof of the part concerning PIC decoding is similar and hence is omitted.

We use Theorem 2 to prove this claim. Let $k \in \{1, \ldots, g\}$, $a_k = [a_{k,1}, \ldots, a_{k,\lambda}] \in \Delta\mathcal{A}_{\mathcal{I}_k} \setminus \{0\}$ and $u \in \mathbb{R}^{|\tilde{\mathcal{I}}_k|}$. We need to show that $X = X_{\mathcal{I}_k}(a_k) + X_{\tilde{\mathcal{I}}_k}(u)$ is of rank $N$. Because $Q$ is a full-diversity rotation matrix, none of the coordinates of $a_k$ is equal to zero, i.e., $a_{k,j} \neq 0$, for $j = 1, \ldots, \lambda$. Let the $k^{th}$ group of symbols be encoded by the $m^{th}$ diagonal layer of the design in (6), i.e., $K'(m-1)+1 \leq k \leq K'm$. Let the weight matrices of the COD $\mathbf{W}$ be $A'_1, \ldots, A'_{K'}$. Then, the matrix $X$ equals

$$\begin{bmatrix}
\mathbf{0} & \mathbf{0} & \cdots & \mathbf{0} \\
\vdots & \vdots & \vdots & \vdots \\
\mathbf{0} & \mathbf{0} & \cdots & \mathbf{0} \\
W(m,1) & \mathbf{0} & \cdots & \mathbf{0} \\
W(m+1,1) & W(m,2) & \cdots & \vdots \\
\vdots & W(m+1,2) & \ddots & \mathbf{0} \\
\vdots & \vdots & \ddots & W(m,L) \\
\vdots & \vdots & \cdots & W(m+1,L) \\
\vdots & \vdots & \cdots & \vdots \\
W(n,1) & \vdots & \cdots & \vdots \\
\mathbf{0} & W(n,2) & \cdots & \vdots \\
\vdots & \vdots & \ddots & \vdots \\
\mathbf{0} & \mathbf{0} & \cdots & W(n,L)
\end{bmatrix}, \quad (13)$$

wherein the first $m-1$ diagonal layers are composed of all-zero matrices. For $p = m, \ldots, n$, and $\ell = 1, \ldots, L$, each $W(p, \ell)$ is some real linear combination of the matrices $A'_1, \ldots, A'_{K'}$.

Now consider an $\ell \in \{1, \ldots, L\}$. We have $W(m, \ell) = \sum_{i=1}^{K'} u_{i,m,\ell} A'_i$ for some choice of real numbers $u_{i,m,\ell}$, $i = 1, \ldots, K'$, that depends upon $a_k$ and $u$. Since the $m^{th}$ layer encodes the $k^{th}$ group of information symbols, there exist $i_0 \in \{1, \ldots, K'\}$ and $j_0 \in \{1, \ldots, \lambda\}$ such that $u_{i_0,m,l} = a_{k,j_0} > 0$. By using the property of CODs that $A'^H_i A'_j + A'^H_j A'_i = 2\delta_{i,j} I_{N'}$ [15], we have

$$det\left(W(m,\ell)^H W(m,\ell)\right) = det\left(\sum_{i=1}^{K'} u_{i,m,\ell}^2 I_{N'}\right)$$

$$= \left(\sum_{i=1}^{K'} u_{i,m,\ell}^2\right)^{N'} \geq u_{i_0,m,\ell}^{2N'} > 0. \quad (14)$$

Thus, $W(m, \ell)$ is full-ranked for each $\ell = 1, \ldots, L$. Given that $W(m,1), W(m,2), \ldots, W(m,L)$ are all full-ranked, it is straightforward to show that the matrix $X$ in (13) is also full-ranked. This completes the proof. ∎

## APPENDIX F
### KNOWN STBCs AS SPECIFIC EXAMPLES OF THE PROPOSED FAMILY OF CODES

When $\mathbf{W}$ in (6) is chosen to be the trivial $1 \times 1$ COD $[s_1 + is_2]$, we get a family of codes first constructed in [9] in the context of point-to-point MIMO channels. This family includes codes for all $N \geq 1$ with parameters $N' = T' = 1$, $L = N$, $1 \leq \lambda \leq N$, $T_2 = N + n - 1$, $g = 2n$ and $K = 2n\lambda$. For $m = 1, \ldots, n$ and $\ell = 1, \ldots, \lambda$, we have $\mathbf{W}(m, \ell) = x_{2(m-1)\lambda+\ell} + ix_{2(m-1)\lambda+\ell+\lambda}$, and no conjugates of these variables appear anywhere in the design in (6). Let $\mathbf{d}_m = [\mathbf{W}_{m,1}^T, \mathbf{W}_{m,2}^T, \ldots, \mathbf{W}_{m,\lambda}^T]^T$, $m = 1, \ldots, n$ and let $z = [\mathbf{d}_1^T \ \mathbf{d}_2^T \cdots \mathbf{d}_\lambda^T]^T$. Since $z$ contains all the complex symbols $x_{2(m-1)\lambda+\ell} + ix_{2(m-1)\lambda+\ell+\lambda}$ that appear in the design (6), there exist matrices $B_j \in \mathbb{C}^{T_2 \times n\lambda}$, $j = 1, \ldots, N$, such that the design (6) equals $[B_1 z \ B_2 z \cdots B_N z]$. Then, the length of the vector $z$ transmitted by the source is $T_1 = n\lambda$. The rate of the DSTBC is $R = \frac{\lambda}{\lambda+1+\frac{N-1}{n}}$. By increasing $n$, rates close to $\frac{\lambda}{\lambda+1}$ can be achieved.

In [9], it was shown that the codes discussed in the previous paragraph include a family of codes in [11] (corresponding to $\mathbf{W} = [s_1 + is_2]$ and $\lambda = N$) along with a lower decoding complexity grouping scheme, and the Toeplitz codes (corresponding to $\mathbf{W} = [s_1 + is_2]$ and $\lambda = 1$) as special cases. In [10], single complex symbol PIC decodable codes were constructed for $N = 2, 4$ antennas. Both codes belong to the class of STBCs constructed in Section IV. With $\mathbf{W} = [s_1 + is_2]$, $n = 2$ and $\lambda = 2$, we get the $N = 2$ code, and with $\mathbf{W}$ as the Alamouti design, $n = 2$ and $\lambda = 2$ we get the $N = 4$ code. A family of codes constructed independently in [9] and [12] is also a special case of the construction given in Section IV. This family of codes corresponds to $\mathbf{W}$ being the Alamouti design and $\lambda = \frac{N}{2}$.

The new class of DSTBCs constructed in Section IV are based on the full-diversity criteria in Theorems 1 and 2. These are identical to the criteria given in [9] for an STBC to give full-diversity in a point-to-point MIMO channel with PIC/PIC-SIC decoding. Hence, the family of DSTBCs constructed in Section IV can be used as full-diversity PIC/PIC-SIC decodable STBCs in point-to-point MIMO channel as well.